\documentclass[%
 reprint,
 amsmath,amssymb,
 aps,
]{revtex4-2}

\usepackage{graphicx} 
\usepackage{amsmath,bm}
\usepackage{placeins}
\usepackage[usenames,dvipsnames]{xcolor}
\usepackage[colorlinks, linkcolor=blue!50!black, urlcolor=blue!50!black, citecolor=blue!50!black]{hyperref}

\newcommand{\Liou}{\mathcal{L}}
\newcommand{\TE}{\mathcal{U}}
\newcommand{\Proj}{\mathcal{P}}
\newcommand{\ProjQ}{\mathcal{Q}}

\newcommand{\ProjM}{\mathcal{P}_M}
\newcommand{\ProjS}{\mathcal{P}_*}
\newcommand{\ProjQS}{\mathcal{Q}_*}

\newcommand{\PS}{{P}^*}

\newcommand{\etaM}{\eta^\text{M}}
\newcommand{\KM}{K^\text{M}}
\newcommand{\OmegaM}{\Omega^\text{M}}

\newcommand{\etaS}{\eta^*}
\newcommand{\tetaS}{\tilde{\eta}^*}
\newcommand{\KS}{K^*}
\newcommand{\OmegaS}{\Omega^*}

\newcommand{\etaI}{\eta^\text{I}}
\newcommand{\KI}{K^\text{I}}

\newcommand{\rhost}{\rho_\text{st}}

\newcommand{\diff}{\text{d}}

\begin{document}

\setlength{\belowdisplayskip}{3pt} \setlength{\belowdisplayshortskip}{3pt}
\setlength{\abovedisplayskip}{3pt} \setlength{\abovedisplayshortskip}{3pt}

\title{ Coarse-graining far-from-equilibrium dynamics using oblique projectors }
\author{Gerhard Jung}
\email[Corresponding author: ]{gerhard.jung.physics@gmail.com}
\affiliation{Institut f\"ur Theoretische Physik, Universit\"at Innsbruck, 6020 Innsbruck, Austria}
\date{\today}

\begin{abstract}
We generalize the Mori-Zwanzig projection operator formalism by introducing an oblique projector using biorthogonal variables to coarse-grain the dynamics of systems far from equilibrium, such as active or living matter. Improving on previously used orthogonal projectors, the oblique projector correctly decomposes the emergent interactions of the coarse-grained particles into systematic and stochastic contributions. We show that the formalism directly connects to established results in non-equilibrium linear response theory and adiabatic perturbation theory. Based on this formalism, we develop a coarse-graining algorithm which we apply to extract the emergent non-equilibrium chiral dynamics of a passive tracer in a chiral active bath. Our theory correctly predicts the odd transport properties and a confinement-induced shift of the chiral frequency of the tracer. 
\end{abstract}

\maketitle

Systems far from equilibrium are abundant in nature, including microorganisms, such as cells \cite{angelini2011glass} or bacteria \cite{aranson2022bacterial}, and macroorganisms, such as schools of fish \cite{lopez2012behavioural} or flocks of birds \cite{cavagna2014bird}. A central objective of active matter research is characterizing and understanding the dynamics of such complex systems \cite{bechinger2016active,gompper20252025}. This often requires a reduction of the underlying microscopic dynamics and the extraction of coarse-grained equations of motion \cite{SCHILLING20221}. The most common ``top-down'' approach is to start from postulating coarse-grained models, and a posteriori validating whether the model reproduces macroscopic properties such as collective behaviour. Examples are active Brownian particles \cite{schimansky1995structure,romanczuk2012active} or the Vicsek model \cite{vicsek1995novel}, which are both now paradigmatic models in soft matter physics. 

The complementary ``bottom-up'' approach is a systematic extraction of coarse-grained models from microscopic systems. The rigorous theoretical foundation of this approach for equilibrium systems is the Zwanzig projection operator formalism \cite{Zwanzig1961,Zwanzig2001}, in particular combined with the linear Mori projector \cite{Mori1965}. Using this Mori-Zwanzig (MZ) formalism allows to derive the generalized Langevin equation (GLE), a stochastic differential equation describing the motion of the coarse-grained variables. The MZ formalism has been used for first-principle theories of liquids and glasses \cite{bengtzelius1984dynamics,janssen2018mode}, and to develop coarse-graining algorithms \cite{hijon2009mori,klippenstein2021introducing} to derive non-Markovian models for liquids \cite{rotenberg2014MZ,straube2020rapid,klippenstein2023bottom}, polymers \cite{li2017computing} and colloids \cite{jung2017iterative,jung2018generalized,shea2022active}. Projection operator formalisms have also been applied to non-stationary Hamiltonian dynamics with time-independent \cite{robertson1966equations,kawasaki1973theory,ochiai1973new,willis1974time,grabert1977microdynamics,grabert1982projection,meyer2017non,Meyer2019Dynamics} or even time-dependent Hamiltonians \cite{Shibata1999_MZ,mcphie2001generalized,te2019mori,izvekov2021mori,netz2024neq,hery2024derivation,izvekov2025mori,hery2026nonequilibrium}. The open question is whether such techniques can also be applied to derive GLEs for systems with intrinsically time-irreversible dynamics \cite{Stella2014GLE,Vandebroek2017,Jung_2022,loos2022stochastic,Jung_2024}, such as active matter \cite{netz2020cell,shea2022active,feng2023active,busiello2024active,shea2024active,netz2024cell,Wijland2025PRE,Wijland2025PRL}.

In this Letter, we therefore apply the projection operator formalism to non-Hamiltonian dynamics. We find that the linear Mori projector does not correctly coarse-grain these far-from-equilibrium systems which we can account to the emergence of non-trivial cross-correlations in non-equilibrium systems violating the equipartition theorem. To overcome this problem we propose an alternative oblique linear projector and demonstrate that it resolves the shortcomings of the Mori projector. We develop a generic numerical algorithm enabling the extraction of complex coarse-grained models, which we apply to the emergent non-equilibrium chiral dynamics of passive tracers in chiral active matter \cite{kalz2022collisions,poggioli2023odd,li2023chirality,kalz2026_reversal, grober2026hydrodynamic,Wijland2025PRE,Wijland2025PRL,goerlich2026particle,pagonabarraga2026chiral1}.

We start from the general equations of motion,
\begin{equation}\label{eq:micro}
    \dot{\bm{v}}(t) = \bm{F}(\bm{v}) + \bm{\xi} \bm{W}(t),
\end{equation}
where $\bm{v} \in \mathbb{R}^{N}$ is the state vector (e.g., positions and velocities), $\bm{F}: \mathbb{R}^{N} \rightarrow \mathbb{R}^{N}$ is a general function (e.g., dissipative or conservative forces), $\bm{\xi}\in \mathbb{R}^{N}\times\mathbb{R}^{N}$ is the noise matrix and $\bm{W}(t) \in \mathbb{R}^{N}$ is Gaussian white noise, $\langle W_i(t) W_j(0) \rangle = \delta_{ij}\delta(t),$ with zero mean. In the following, we will denote Eq.~(\ref{eq:micro}) as the ``microscopic'' dynamics, although this description may also correspond to some basis level of coarse-graining, such as Langevin dynamics. The time evolution can be written using a stochastic operator (Einstein notation) \cite{Li2020overdamped,Szamel2021_proj},
\begin{alignat}{2}
   \Liou(t) &=  \left(F_i + \frac{1}{2}  \xi_{ij} \frac{\diff}{\diff v_j}\right)\frac{\diff}{\diff v_i} &+& \xi_{ij} W_j(t)\frac{\diff}{\diff v_i} \nonumber\\
   &= \Liou_F &+& \Liou_\xi(t), 
\end{alignat}
where we have separated the time-independent contributions $\Liou_F$ from the time-dependent contributions $\Liou_\xi(t).$ Using the negatively time-ordered exponential, we define the evolution operator starting at time $\tau < t$ \cite{j2007statistical,Koch2024_force}, 
\begin{equation}
    \TE_\Liou(t,\tau) = \exp_-\left( \int_{\tau}^t \diff s \Liou(s) \right),
\end{equation}
and rewrite the equations of motion,
\begin{align}
  \dot{\bm{v}}(t) = \TE_\Liou(t,\tau) \Liou(t)  \bm{v}(\tau).
\end{align}
The following derivation will be kept brief using the component $v_0$ as the one-dimensional coarse-grained variable, which could describe the velocity of a tracer (see \emph{End Matter} for a general multi-dimensional derivation). We introduce the general, linear projection operator $\Proj A = \langle P(A,\bm{v}(\tau)) \rangle  v_0(\tau)$, acting on an observable $A.$ Specifying the function $P(A,\bm{v}):\mathbb{R} \times \mathbb{R}^{N} \rightarrow \mathbb{R}$ then defines the projector. For example, the one-dimensional Mori projector $\ProjM$ is defined by $P^M(A,\bm{v}) =  Av_0  \langle v_0^2 \rangle^{-1}$ \cite{Mori1965,Zwanzig2001}. We now follow the lines of Refs.~\cite{Koch2024_force,Koch2025_phd} to derive the GLE,
\begin{align}\label{eq:GLE_general}
    \dot{v}_0(t) &= \Omega v_0(t) - \int_\tau^t \diff s K(t-s) v_0(s) + \eta(t),
\end{align}
where the formal expressions for $\Omega$, $K(t)$ and $\eta(t)$ are given in the \emph{End Matter}. Here, $F^D(t) = \Omega v_0(t) - \int_\tau^t \diff s K(t-s) v_0(s)$ describes the systematic forces in the GLE. $\eta(t)$ is identified as the fluctuating force and usually modeled as a stochastic process in coarse-grained models. Similarly, we can start from \cite{zhu2022_MZ}
\begin{align}\label{eq:Volterra_start}
  \frac{d}{dt} \Proj v_0(t) &= \Proj  \TE_\Liou(t,\tau) \Liou(t)  v_0(\tau)
\end{align}
to derive the deterministic Volterra equation
\begin{equation}\label{eq:Volterra}
  \dot{C}(t) = \Omega C(t) - \int_0^t \diff s K(t-s) C(s).
\end{equation}
with $C(t)=\langle P(v_0(t+\tau),\bm{v}(\tau)) \rangle$. We have assumed that the dynamics is stationary and thus $C(t)$ does not depend on $\tau$. If we insert the Mori projector into Eq.~(\ref{eq:Volterra}) we find the  Volterra equation for the Mori kernel $\KM(t)$ \cite{Zwanzig2001,SHIN2010316},
\begin{equation}\label{eq:Volterra_Mori}
  \dot{C}^V(t) = \OmegaM C^V(t) - \int_0^t \diff s \KM(t-s) C^V(s),
\end{equation}
involving the velocity autocorrelation function (VACF), $C^V(t) = \langle v_0(t+\tau) v_0(\tau) \rangle.$ In many coarse-graining procedures, Eq.~(\ref{eq:Volterra_Mori}) is used to extract memory kernels from microscopically extracted correlation functions \cite{SHIN2010316,klippenstein2021introducing}.

In the following, we illustrate why Eq.~(\ref{eq:Volterra_Mori}) cannot generally hold far from equilibrium. We study the non-reciprocal model \cite{Doerries_2021,Jung_2022,Jung_2024}, as special case of Eq.~(\ref{eq:micro}),
	\begin{align}
	\dot{v}_0(t) &= - 2 v_0(t) + v_1(t) \label{eq:non_rec1}\\
	 \dot{v}_1(t) &= - v_1(t) + v_0(t) +  W_1(t) \label{eq:non_rec2}.
	\end{align}
    By integrating out the dynamics for $v_1(t)$ and inserting it into the first equation (see \emph{End Matter}), we find analytically in the stationary limit the following Volterra equation
    \begin{align}\label{eq:Volterra_non_rec}
    	\dot{C}^V(t) &= - 2 {C}^V(t)  - \int_{0}^{t} \hspace{-0.1cm} \text{d}s\KI(t-s) C^V(s) + \langle v_0 v_1 \rangle e^{- t}.
	\end{align}
    Eq.~(\ref{eq:Volterra_non_rec}) is clearly distinct from the Mori Volterra Eq.~(\ref{eq:Volterra_Mori}) since $\langle v_0 v_1 \rangle = 1/3 \neq 0.$ Therefore, $\KM(t)$ will not correspond to the analytically derived kernel $\KI(t)$ and the coarse-grained systematic force is therefore not correctly identified by the MZ formalism \cite{Jung_2022,Jung_2024}. As a direct consequence, the fluctuating force $\etaM(t)$ must contain this missing systematic contribution and can thus not be interpreted as a stochastic process.

\begin{figure}
    \centering
    \includegraphics[width=1.04\linewidth]{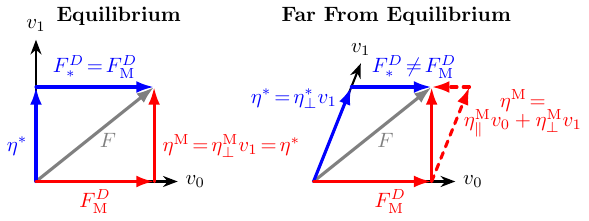}
    \caption{Vectorial illustration of how cross-correlations $\langle v_0 v_1 \rangle \neq 0$ impact the force decomposition ${F} = {F}^D + {\eta} $ for the Mori projector, $\ProjM$, and the newly proposed oblique projector, $\ProjS$. Far from equilibrium ${\etaM}$ depends on the basis vector ${v}_0$, as indicated by the dashed arrows, and can thus not be interpreted as a fluctuating force.}
    \label{fig:Illustration}
\end{figure}

The crucial problem is to find a projection operator $\ProjS$ which overcomes this major shortcoming of the Mori projector $\ProjM$. The difficulties emerge due to the cross-correlations $\langle v_0 v_1 \rangle $ which are absent in equilibrium dynamics. These cross-correlations can be illustrated as non-orthogonal basis vectors in a two-dimensional Euclidean space where correlations correspond to parallel components (see Fig.~\ref{fig:Illustration}). The MZ formalism decomposes the total force $F$ into a component correlated with ${v}_0$ which represents the systematic force $F_\text{M}^D,$ and the fluctuating force $\etaM$. In equilibrium, this fluctuating force can be interpreted as a stochastic process since its expansion is independent of the coarse-grained variable ${v}_0$ (see Fig.~\ref{fig:Illustration}, left panel). Far from equilibrium, the Mori projector still enforces that $\etaM$ is \emph{orthogonal} to ${v}_0$. However, its expansion in the non-orthogonal basis $({v}_0,{v}_1)$ nevertheless \emph{depends} on the coarse-grained variable ${v}_0$ (see dashed arrows in Fig.~\ref{fig:Illustration}). Interpreting $\etaM$ as a stochastic process is therefore incorrect since it depends systematically on the coarse-grained variable ${v}_0$ with $\etaM_\parallel\neq 0$. We can thus conclude that far-from-equilibrium dynamics requires going beyond the Mori projector and defining a fluctuating force $\etaS$ which may be \emph{non-orthogonal} to $\bm{v}_0$, however, its expansion must be \emph{independent} of $\bm{v}_0.$  This new concept is the first main result of this Letter. The remainder of this manuscript demonstrates how this concept can be realized using a linear projector and how it can be applied to systematically construct coarse-grained models far from equilibrium.

To implement this concept we define an oblique projection operator $\ProjS A = \langle A v_* \rangle v_0$, where $v_*$ is the biorthogonal coarse-grained variable which fulfills $\langle v_i v_* \rangle = \delta_{i0}$.  For general systems this can be realized by $P^*(A,\bm{v}) = A  \partial_{v_0} \ln \rhost(\bm{v}),$ where $\rhost(\bm{v})$ is the stationary probability distribution of the microscopic system. This definition is equivalent to $\PS(A,\bm{v}) = \partial_{v_0}A$, which implies that for any linear observable $A = a_0 v_0 + a_1 v_1$ we have $\ProjS A = a_0 v_0$ and thus $\ProjQS A = (1 -  \ProjS )A= a_1 v_1$. In consequence, $\etaS$, which purely lives in the $\ProjQS-$space, is independent of $v_0.$ From a physical perspective, different from the Mori projector which is based on correlations of $A$ with $v_0$, the oblique operator is thus based on the response of $A$ to changes in $v_0.$  Such oblique projections are known in signal processing to separate data from noise \cite{behrens1994signal}, and in quantum mechanics \cite{brody2014biorthogonal}, but they have never been used in the context of the MZ formalism.

Inserting the projector $\ProjS$ into the general GLE (\ref{eq:GLE_general}) we find the oblique GLE,
\begin{equation}\label{eq:GLE_dual}
    \dot{v}_0(t) = \OmegaS v_0(t) - \int_\tau^t \diff s \KS(t-s) v_0(s) + \etaS(t) 
\end{equation}
with the formal expressions for $\OmegaS$, the oblique kernel $\KS$ and the fluctuating force $\etaS$ given in the \emph{End Matter}.
Additionally, if we insert $\ProjS$ into Eq.~(\ref{eq:Volterra}) we can derive the Volterra equation
\begin{equation}\label{eq:Volterra_dual}
  \dot\chi(t) = \OmegaS \chi(t) - \int_0^t \diff s \KS(t-s) \chi(s)
\end{equation}
for the response function 
\begin{equation}\label{eq:response_dual}
\chi(t) = \langle v_0(t)  \partial_{v_0} \ln \rho_\text{st}(\bm{v}) \rangle =  \langle \partial_{v_0} v_0(t)  \rangle.
\end{equation}
This is the second main result showing generally that coarse-graining far from equilibrium requires using Volterra equations based on response instead of correlation functions \cite{netz2018neq}. In equilibrium, both equations coincide due to the first fluctuation-dissipation relation (1FDR) \cite{kubo1966fluctuation}. 

Importantly, the oblique projector bridges the gap between projection operator techniques and known results in non-equilibrium statistical physics. The above relation for the response function $\chi(t) $ in Eq.~(\ref{eq:response_dual}) is known as a non-equilibrium fluctuation-dissipation relation which can be derived using non-equilibrium linear response theory \cite{MARCONI2008111,Baiesi_2013,Vulpiani2019_fdr,Caprini_2021,caprini2021_fdr} (see SM \cite{suppmat} for an alternative derivation). Additionally, the oblique memory kernel $\KS = -\left\langle \Liou_F \etaS(t) \partial_{v_0} \ln \rhost(\bm{v})   \right\rangle $  is similar to results obtained from adiabatic perturbation theory and often called an Agarwal-like formula \cite{Agarwal1972,Solon_2022,Tailleur2022_agarwal,Wijland2025PRE,Wijland2025PRL}. Finally, if we use the oblique projector $\ProjS$ to coarse-grain the non-reciprocal model Eqs.~(\ref{eq:non_rec1},\ref{eq:non_rec2}) we can explicitly evaluate the formal relations for $\KS$ and $\etaS$ to find the Volterra equation Eq.~(\ref{eq:Volterra_non_rec}) (see \emph{End Matter}). For the non-reciprocal model, the oblique GLE (\ref{eq:GLE_dual}) is therefore equivalent to the analytically derived GLE, thus resolving the issue of the Mori projector identified at the beginning of this Letter. 

We can use the oblique projector to propose a coarse-graining algorithm far from equilibrium from microscopic simulations or experiments. First, we extract the response functions $\left\langle \partial_{v_0} F_0 \right\rangle$  and $\chi(t)$. Second, we determine $\OmegaS$ and invert Eq.~(\ref{eq:Volterra_dual}) to reconstruct the memory kernel $\KS$ from $\chi(t)$. Third, assuming stationary dynamics with $\tau \rightarrow -\infty$ we use the recorded unperturbed trajectories from microscopic simulations to extract the fluctuating force $\etaS(t)$ which is the only unknown quantity in the GLE~(\ref{eq:GLE_dual}). Numerical details of this algorithm are presented in the SM \cite{suppmat}.

The Mori and the oblique projector are applied and compared by extracting the equations of motion describing the emergent chiral dynamics of a passive tracer in a chiral active bath in two dimensions. We study the same system as introduced in Refs.~\cite{Wijland2025PRE,Wijland2025PRL}. The bath particles are modeled as chiral active Brownian particles with propulsion force $f_0$, friction constant $\gamma,$ rotational diffusion constant $D_r,$ persistence length $\ell_p = f_0 / D_r \gamma = 10$ and an additional chiral frequency $\omega$. This induces circular trajectories with typical gyroradius $ {\ell}_g = f_0 / |\omega| \gamma.$  The passive tracer is an inertial particle with mass $m$.  Both the tracer and the chiral particles are spherical and interact solely via a conservative, purely repulsive force. Different from Ref.~\cite{Wijland2025PRL} we add translational noise and replace the hard-core repulsive potential by a softer linear repulsion, which smoothens the memory kernel and thus simplifies the numerical reconstruction. The typical length scale of the repulsion is $r = 0.25 \ell_p $ (see SM for model details  \cite{suppmat}). We extract the response functions from microscopic simulations using standard methods as described in Refs.~\cite{jung2021fdr,jung2026_nonlinear}. After applying the coarse-graining algorithm, we map the reconstructed Mori and oblique GLE onto an extended Markov model with the same statistical properties as the reconstructed GLEs. This allows us to perform very efficient coarse-grained simulations of the passive tracer, by numerically solving this extended Markov model (see SM \cite{suppmat}). 

\begin{figure}
    \centering
    \includegraphics[width=1.04\linewidth]{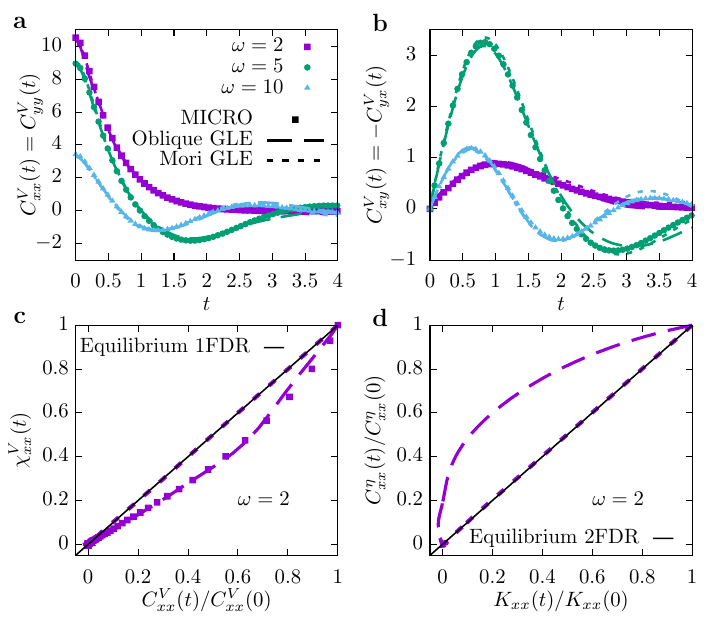}
    \caption{Correlation functions and fluctuation-dissipation relations of a passive tracer in chiral active matter, comparing the microscopic simulations (MICRO), the oblique GLE and the Mori GLE. \textbf{(a)} Diagonal component of the VACF for different chiral frequencies $\omega$. \textbf{(b)} Same as (a) for the off-diagonal component. \textbf{(c)} Relation between the diagonal components of the VACF and the response function $\chi^V(t)$. \textbf{(d)} Relation between the memory kernel $K(t)$ and $C^\eta_{\alpha \beta}(t)=\langle \eta_\alpha(t) \eta_\beta(0)\rangle.$  }
    \label{fig:VACF}
\end{figure}

We compare the coarse-grained dynamics of the tracer with the microscopic dynamics. We find a very good agreement between the microscopic VACF $C^V_{\alpha \beta}(t) = \langle v_{0,\alpha}(t) v_{0,\beta} \rangle,$ and the Mori and oblique GLE (see Fig.~\ref{fig:VACF}a,b). Small differences can be accounted to numerical errors in the sampling, the memory reconstruction and the Markovian embedding. Importantly, we observe that the off-diagonal terms $C^V_{xy}(t)$ are non-zero and anti-symmetric and thus reveal a chiral motion of the tracer induced by the active bath. To further investigate the non-equilibrium nature of the dynamics we investigate the fluctuations-dissipation relations using the same representation as in Ref.~\cite{Vulpiani2019_fdr}. The 1FDR, i.e., the relationship between response and correlation functions, is clearly violated in the microscopic system \cite{Fuchs2009_fdr,maggi2017memory} (see Fig.~\ref{fig:VACF}c). We observe a qualitative difference between the Mori and the oblique GLE. While the former perfectly follows the equilibrium 1FDR and thus indicates that the non-equilibrium dynamics is mapped onto an effective equilibrium model \cite{jung2021fdr,shea2022active,shea2024active}, the oblique GLE violates the 1FDR, in very good agreement with the microscopic simulations. We also find that the oblique GLE predicts a violation of the second fluctuation-dissipation relation (2FDR) \cite{Zamponi_2005,maes2013fdr,Cui2018GLE,zaccone2023fdr}, i.e., the relationship between the memory kernel and the fluctuating force (see Fig.~\ref{fig:VACF}d). 

\begin{figure}
    \centering
    \includegraphics[width=0.99\linewidth]{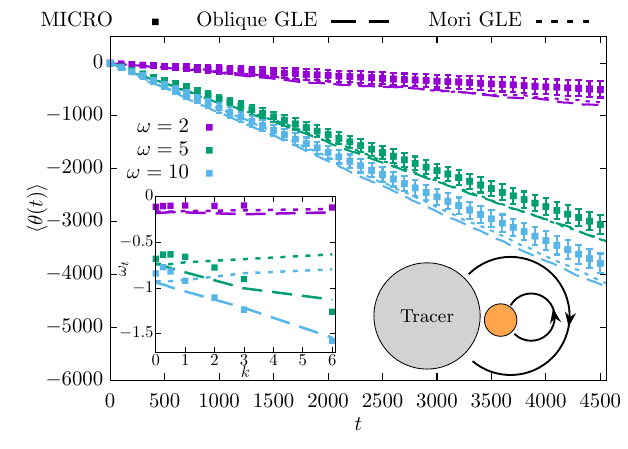}
    \caption{Mean angular drift $\langle \theta(t) \rangle$ of the passive tracer. The data is averaged over many different trajectories, error bars show standard deviation. The illustration highlights opposite rotation of the tracer and the chiral particles. The inset shows the emergent chiral frequency of the passive tracer $\omega_t = d \langle \theta(t) \rangle / dt $ when confined in a harmonic potential of strength $k.$ }
    \label{fig:harmonic}
\end{figure}

We can extract information on the emergent angular drift of the passive tracer. For this we calculate the smoothed velocity vector $\tilde{\bm{v}}(t) = \int_{t-0.25}^t ds\, \bm{v}_0(s)$ and thus define the angle $ \theta(t) = \text{atan2}(\tilde {v}_y(t), \tilde {v}_x(t) ) $ \cite{goerlich2026particle}. We observe that the passive tracer indeed inherits a pronounced angular drift from the chiral particles (see Fig.~\ref{fig:harmonic}). Remarkably, for all $\omega$ we find a pronounced clockwise rotation in the opposite direction of the chiral particles. If we additionally confine the tracer by applying an external harmonic force with strength $k$, the Mori GLE predicts a slight decrease of the absolute value of the emergent chiral frequency  $\omega_t = d \langle \theta(t) \rangle / dt $, while the oblique GLE predicts a strong increase (see Fig.~\ref{fig:harmonic}, inset). The prediction of the oblique GLE is clearly confirmed by microscopic simulations, which shows that the oblique GLE can be used to make non-trivial predictions for the non-equilibrium dynamics, inaccessible by the MZ formalism, which is another major result of this Letter.

\begin{figure}
    \centering
    \includegraphics[width=0.99\linewidth]{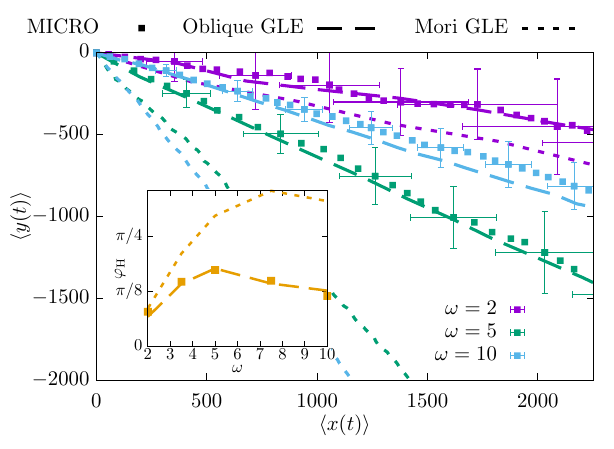}
    \caption{Odd transport of the passive tracer with an external pulling force $\bm{F}_\text{ext} = [F_x,0]$. The data is averaged over many different trajectories, error bars show standard deviation. The inset shows the hall angle $\tan (\varphi_\text{H}) = -y(t\rightarrow \infty) / x(t\rightarrow \infty) $ for various chiral frequencies $\omega.$}
    \label{fig:extForce}
\end{figure}

Finally, we investigate the emergent odd transport of the tracer \cite{Wijland2025PRL,goerlich2026particle}. We observe that a constant force in $x-$direction leads to a strong systematic drift in negative $y-$direction (see Fig.~\ref{fig:extForce}). While the Mori GLE, however, massively overestimates odd transport, the oblique GLE is in good agreement with the microscopic simulations. In analogy to the Hall effect where external magnetic fields induce currents transverse to the direction of an electric field, we can define the hall angle $\varphi_\text{H}$ describing the relative amplitude of the odd transport. We find that $\varphi_\text{H}$ significantly increases until $\omega \approx 5$ and then decreases upon further increase of $\omega.$ The hall angle is thus maximized when the gyroradius ${\ell}_g(\omega=5) = 2.0\sigma \approx r$ corresponds to the length scale of the active-passive interactions. For significantly larger $\omega$ the chiral particles spin in very small circles and the transfer of its chiral properties to the tracer becomes inefficient.

Compared to a recent experimental study of a passive object interacting with self-propelled chiral particles \cite{goerlich2026particle}, the observed mean angular drift and odd transport has a different sign. This discrepancy has been rationalized with the presence of dry friction \cite{Antonov2024_dry} of their passive object (see Fig.~4c,d in \cite{goerlich2026particle}) which is not present in our model. Generally, it is known that the sign of tracer drift and odd transport is non-universal and depends qualitatively on the specific model \cite{kalz2026_reversal}.

In conclusion, we have introduced an oblique projector which allows to derive dynamically-consistent GLEs for systems far from equilibrium. While previous approaches defined the fluctuating force via an orthogonality condition, the oblique projector ensures that the fluctuating force is independent of the coarse-grained variables and can thus be interpreted as a stochastic process. The applicability of the theory is illustrated on the example of emergent chiral dynamics of a passive tracer in a chiral active bath. We can show that the oblique GLE correctly reproduces the non-equilibrium nature of the microscopic system and can be used to make non-trivial predictions on emergent persistent rotations in harmonic confinement. The new formalism therefore represents an essential step towards accurately coarse-graining far-from-equilibrium dynamics. It should be emphasized that the oblique GLE for the tracer cannot be replaced by a simple Langevin equation using the Markovian approximation, since the fluctuating force correlation matrix would be anti-symmetric. This highlights the relevance of using a non-Markovian description to simulate odd diffusive particles \cite{loewen2026_odd} \footnote{Previous works have simulated odd diffusive particles using a Langevin equation with a diagonal noise matrix. It is not clear whether this could correspond to the outcome of a coarse-graining procedure \cite{kalz2022collisions,loewen2026_odd}. }.  In future research it would be interesting to extend the formalism to overdamped \cite{Li2020overdamped} and non-stationary dynamics \cite{meyer2017non} or dense active matter \cite{liluashvili2017mode,Szamel2019_glass,janssen2019active}. The long-time goal is also to apply the reconstruction algorithm to experiments of colloids \cite{bechiner2020_trap}, microswimmers \cite{tung2017fluid} or macroscopic active matter \cite{goerlich2026particle}.  More complex systems may potentially require data-driven techniques \cite{vroylandt2022memory,Lei2025ml} to extract the oblique GLE directly from microscopic trajectories.

We thank T. Franosch, A. Lüders, R. Goerlich and A. Antonov for very helpful discussions. This research was funded by the Austrian Science Fund (FWF) 10.55776/PAT1139125.

\bibliography{library}

\section*{End Matter}

\emph{Multi-dimensional projection operator formalism} -- We start by introducing the $M-$dimensional observable, $\bm{A}(\bm{v}(\tau)):\mathbb{R}^N \rightarrow \mathbb{R}^M,$ for which we can write the time-evolution,
\begin{align}
  \bm{A}(t;\bm{v}(\tau)) &= \TE_\Liou(t,\tau) \bm{A}(\bm{v}(\tau)),\\
  \frac{\diff}{\diff t} \bm{A}(t;\bm{v}(\tau)) &= \TE_\Liou(t,\tau) \Liou(t)  \bm{A}(\bm{v}(\tau)).
\end{align}
In the following, we will use the short notations $\bm{A}(\bm{v}(\tau))=\bm{A}$, and  $\bm{A}(t;\bm{v}(\tau)) = \bm{A}(t)$. The observable $\bm{A}$ will denote the coarse-grained variables in the following. We now introduce the multi-dimensional projection operator, $\Proj {B}_\alpha = \langle P_{\alpha \beta}({B}_\alpha,\bm{A}) \rangle  A_\beta$, acting on an observable $\bm{B}$. Within this notation, the Mori projector can be defined as $P^M_{\alpha \beta}({B}_\alpha,\bm{A}) = {B}_\alpha A_\gamma G_{\gamma\beta}^{-1},$ with $G_{\alpha \beta} = \langle A_\alpha A_\beta \rangle.$ We can use the projector to expand the time evolution of the coarse-grained variables,
\begin{align}\label{eq:em:timeevo}
  \frac{\diff}{\diff t} \bm{A}(t) = \TE_\Liou(t,\tau) \Big( \Proj \Liou_F  +\ProjQ \Liou_F  +  \Liou_\xi(t)  \Big) \bm{A}.
\end{align}
As is commonly done in projection operator formalisms, we then use the Dyson relation to rewrite the time-ordered time-evolution operator \cite{Koch2024_force,Koch2025_phd},
\begin{align}\label{eq:em:dyson}
  \TE_\Liou(t,\tau) \ProjQ &= \ProjQ \mathcal{G}(t) + \int_\tau^t \diff s \, \TE_\Liou(s,\tau) (\Proj \Liou_F + \Liou_\xi(s)) \ProjQ \mathcal{G}(t-s) 
\end{align}
with $\mathcal{G}(t)=\exp(t \Liou_F \ProjQ )$. Combining Eqs.~(\ref{eq:em:timeevo}) and (\ref{eq:em:dyson}), we find the GLE,
\begin{align}
    \frac{\diff \bm{A}(t)}{\diff t} &= \bm{\Omega} \bm{A}(t) - \int_\tau^t \diff s \bm{K}(t-s) \bm{A}(s) + \bm{\eta}(t)
\end{align}
where
\begin{align}
    {\Omega}_{\alpha \beta} &= \langle P_{\alpha\beta}(\Liou_F A_\alpha, \bm{A}) \rangle\\
    \tilde \eta_\alpha(t) &= \ProjQ \mathcal{G}(t) \Liou_F A_\alpha\\
    {K}_{\alpha \beta}(t) &=  -\langle P_{\alpha\beta}(\Liou_F \tilde \eta_\alpha(t),A_\beta)  \rangle\\
     \eta_\alpha(t) &= \tilde \eta_\alpha(t) + \TE_\Liou(t,\tau)  \Liou_\xi(t) A_\alpha \nonumber\\
     &\quad+ \int_\tau^t \diff s\,  \TE(s,\tau) \Liou_\xi(s) \ProjQ \mathcal{G}(t-s) \Liou_F A_\alpha.
\end{align}
We can also rewrite the expression for the memory kernel ${K}_{\alpha \beta}(t) =  -\langle P_{\alpha\beta}(\Liou_F \eta_\alpha(t),A_\beta)  \rangle$ because the microscopic noise $W_j(t)$ is orthogonal to the coarse-grained variables.
In particular, for $M=1$ and $A=v_0$ we arrive at the GLE (\ref{eq:GLE_general}) with the formal expressions,
\begin{align}
    \Omega &= \langle P(\Liou_F v_0,v_0) \rangle \label{eq:Omega_general}\\
            \tilde \eta(t) &= \ProjQ \exp(t \Liou_F \ProjQ ) \Liou_F v_0 \label{eq:eta_general}\\
    K(t) &=  -\langle P(\Liou_F \tilde\eta(t),v_0)  \rangle\\
     \eta(t) &= \tilde \eta(t) + \xi_{0j}W_j(t) + \int_\tau^t \diff s\,  \TE_\Liou(s,\tau) \Liou_\xi(s) \tilde \eta(t-s). \label{eq:eta_general}
\end{align}
We can infer that $\tilde \eta(t)$ lives in the orthogonal $\ProjQ$-space and the remaining terms explicitly include the microscopic noise $W_j(t)$. This is the reason why $\eta(t)$ can usually be modeled as a stochastic process. Inserting the oblique projector, we find explicitly the oblique GLE (\ref{eq:GLE_dual}) with the following relations,
\begin{align}
    \OmegaS &= \left\langle \partial_{v_0} F_0 \right\rangle \label{eq:dual_omega}\\
        \tetaS(t) &= \ProjQS \exp(t \Liou_F \ProjQS ) \Liou_F v_0\\
    \KS(t) &=  -\left\langle \Liou_F \tetaS(t) \partial_{v_0} \ln \rhost(\bm{v})   \right\rangle \label{eq:dual_K}\\
     \etaS(t) &= \tetaS(t) + \xi_{0j}W_j(t) + \int_\tau^t \diff s \, \TE_\Liou(s,\tau) \Liou_\xi(s) \tetaS(t-s) \label{eq:dual_eta}
\end{align}

When we apply the projection operator before taking the time-derivative, we have,
\begin{align}
  \frac{\diff}{\diff t} \Proj \bm{A}(t) &= \Proj  \TE_\Liou(t) \Liou(t)  \bm{A} 
  \\ &= \Proj  \TE_\Liou(t) \left( \Proj \Liou_F +   \ProjQ \Liou_F +  \Liou_\xi(t) \right) \bm{A} .
\end{align}
If we insert the Dyson relation, Eq.~(\ref{eq:em:dyson}), we find the Volterra equation,
\begin{equation}\label{eq:Volterra_EM}
  \frac{\diff}{\diff t} \bm{C}(t,\tau) = \bm{\Omega} \bm{C}(t,\tau) - \int_\tau^t \diff s \bm{K}(t-s) \bm{C}(s,\tau),
\end{equation}
with $C_{\alpha \beta}(t,\tau)=\langle P_{\alpha \beta}(A_\alpha(t),\bm{A}(\tau) \rangle.$ Assuming that the dynamics is stationary, $C_{\alpha \beta}(t,\tau) = C_{\alpha \beta}(t-\tau)$, we can thus derive for $M=1$ and $A=v_0$ the Volterra Eq.~(\ref{eq:Volterra}). Any contribution from the fluctuating force $\eta(t)$ in Eq.~(\ref{eq:Volterra}) vanishes since $\Proj \ProjQ = 0$, per definition, and because the ensemble average of any term containing $\bm{W}(t)$ linearly is zero since it is statistically independent of all the other variables at starting time $\tau< t.$

\emph{Properties of the oblique projector} -- In multiple dimensions, the oblique projector is defined by $\PS_{\alpha \beta}(B_\alpha,\bm{A}) = B_\alpha  \partial_{A_\beta} \ln \rhost(\bm{v}).$ Therefore, we have,
\begin{align}
    \ProjS B_\alpha &= \left[\int \diff\bm{v}^\prime \rhost(\bm{v}^\prime) B_\alpha(\bm{v}^\prime)  \partial_{A_\beta} \ln \rhost(\bm{v}^\prime) \right] A_\beta(\bm{v}),\\
    &=\left[\int \diff\bm{v}^\prime B_\alpha(\bm{v}^\prime)  \partial_{A_\beta} \rhost(\bm{v}^\prime)\right] A_\beta(\bm{v})
\end{align}
If we choose $A_\alpha = v_\alpha,$ we can thus explicitly evaluate,
\begin{align}
    \ProjS B_\alpha &= \left[\int \diff\bm{v}^\prime  \rhost(\bm{v}^\prime) \partial_{v^\prime_\beta} B_\alpha(\bm{v}^\prime)   \right] v_\beta,\\
    &= \langle \partial_{v_\beta} B_\alpha(\bm{v}) \rangle  v_\beta
\end{align}
To better understand the properties of the oblique projector, we can also assume that $\rhost(\bm{v})$ is Gaussian, and explicitly evaluate the Gaussian integrals to derive the relation,
\begin{equation}\label{eq:dual_proj_Gaussian}
\ProjS  B_\alpha = \sum_{\gamma=0}^{N-1} \langle B_\alpha v_\gamma \rangle G^{-1}_{ \gamma \beta}  v_\beta
\end{equation}
While this expression looks identical to the Mori projector, it is very important to note that the sum in Eq.~(\ref{eq:dual_proj_Gaussian}) runs over all the $N$ state variables of the microscopic system and not only over the $M$ coarse-grained variables as in the Mori projector. In particular, if $N=2$ and $M=1$ we have,
\begin{align}
    \ProjS  B &= \frac{\langle B v_0 \rangle \langle v_1^2\rangle-\langle B v_1 \rangle \langle v_0 v_1\rangle}{\langle v_0^2 \rangle \langle v_1^2\rangle- \langle v_0 v_1\rangle^2} v_0,\\
    \ProjM  B &= \frac{\langle B v_0 \rangle \langle v_1^2\rangle}{\langle v_0^2 \rangle } v_0
\end{align}
Therefore, the two projectors coincide if and only if $\langle v_0 v_1\rangle = 0.$ In all other cases, $\ProjS$ differs from $\ProjM$ by explicitly removing any contributions to the coarse-grained variables emerging from correlations with $v_1$.

\emph{Two-dimensional non-reciprocal model} -- Formally integrating the first-order inhomogeneous differential equation (\ref{eq:non_rec2}) and inserting it into Eq.~(\ref{eq:non_rec1}) yields the exact GLE \cite{Jung_2022,Jung_2024},
\begin{equation}\label{eq:GLE_integrate}
    	\dot{v}_0(t) =- 2 v_0(t) - \int_{0}^{t} \text{d}s \KI(t-s) v_0(s) + \etaI(t),
\end{equation}
with memory kernel,
	\begin{align}
	\label{eq:memory_integrate}
	\KI(t) &= -  \exp (- t),
	\end{align}
	and noise,
	  \begin{align}
	\label{eq:noise_integrate_full}
	\etaI(t)&= \sqrt{2 k_B T }  \int_{0}^t \text{d}t^\prime \exp (- t^\prime) W_1(t^\prime) + e^{-t}v_1(0).
	\end{align}
    If we now multiply Eq.~(\ref{eq:GLE_integrate}) by $v_0(0)$ and take the ensemble average, we immediately find Eq.~(\ref{eq:Volterra_non_rec}).

    Following the lines of Refs.~\cite{Jung_2022,Jung_2024} we can evaluate the projection operator formalism. First, we evaluate the direct friction coefficients,
    \begin{align}
        \OmegaM = \langle \dot{v}_0(0) v_0(0) \rangle = 0,\\
        \OmegaS = \langle \partial_{v_0} \dot{v}_0(0) \rangle = -2,
    \end{align}
    where $\langle \dot{v}_0(0) v_0(0) \rangle = 0$ follows immediately from Eq.~(6) in Ref.~\cite{Jung_2022}. Subsequently, we take the time-derivative of Eq.~(\ref{eq:eta_general}) to find,
    \begin{equation}\label{eq:tilde_eta_evo}
        \frac{\diff}{\diff t} \tilde{\eta}(t) = \ProjQ \Liou_F \tilde{\eta}(t).
    \end{equation}
    Due to the linearity of the projector and the time-evolution operator $\Liou_F$ we can evaluate this expression analytically using the ansatz,
    \begin{equation}
        \tilde{\eta}(t) = \sigma_0(t) v_0(0) + \sigma_1(t) v_1(0).
    \end{equation}
    Inserting this ansatz into Eq.~(\ref{eq:tilde_eta_evo}) we can extract the equations of motion for $\sigma_i(t)$ following Ref.~\cite{Jung_2022}. We have for the Mori projector,
    \begin{align}
        \frac{\diff}{\diff t } \sigma_{\text{M},0}(t) &= - 2 \sigma_{0,\text{M}}(t) + 2 \sigma_{1,\text{M}}(t),\\
        \frac{\diff}{\diff t } \sigma_{\text{M},1}(t) &= -  \sigma_{1,\text{M}}(t) +  \sigma_{0,\text{M}}(t)
    \end{align}
    and for the oblique projector,
        \begin{align}
        \sigma_{\text{M},0}(t) &= 0,\\
        \frac{\diff}{\diff t } \sigma_{\text{M},1}(t) &= -  \sigma_{1,\text{M}}(t).
    \end{align}
    As a consequence, we find $ \tetaS(t) = \exp(-t)v_1(0).$ Multiplying the oblique GLE Eq.~(\ref{eq:GLE_dual}) with $v_0$ and taking the stationary average thus immediately yields the Volterra Eq.~(\ref{eq:Volterra_non_rec}). Additionally, inserting this expression into the relation for the oblique memory kernel, we find,
    \begin{align}
        \KS (t) &= - \langle \partial_{v_0} \Liou_F \tetaS(t) \rangle ,\nonumber\\
        &= - \exp(-t) \langle \partial_{v_0} \Liou_F v_1(0) \rangle ,\nonumber\\
        &= - \exp(-t) \langle \partial_{v_0} (- v_1(0) + v_0(0)) \rangle, \nonumber\\
        &= - \exp(-t).
    \end{align}
    As a consequence, we have derived that the oblique GLE (\ref{eq:GLE_dual}) is identical to the exact GLE (\ref{eq:GLE_integrate}). This contrasts results from the Mori GLE \cite{Jung_2022} which features very different equations of motion. In particular, the Mori GLE fulfills the 2FDR, $\langle \etaM(t) \etaM(t) = \langle v_0(0)^2 \rangle \KM(t)$ while the oblique GLE and analytical GLE do not, as can be immediately inferred from the negative memory kernel.
\end{document}


\setlength{\belowdisplayskip}{3pt} \setlength{\belowdisplayshortskip}{3pt}
	\setlength{\abovedisplayskip}{3pt} \setlength{\abovedisplayshortskip}{3pt}
	
	\title{Supplemental Material for ``Coarse-graining far-from-equilibrium dynamics using oblique projectors''}
	
	\author{Gerhard Jung}

	\date{\today}
	
	\maketitle
	
	\setcounter{equation}{0}
	\setcounter{figure}{0}
	\setcounter{table}{0}
	\setcounter{page}{1}
	\renewcommand{\theequation}{S\arabic{equation}}
	\renewcommand{\thefigure}{S\arabic{figure}}
	\renewcommand{\bibnumfmt}[1]{[S#1]}
	\renewcommand{\citenumfont}[1]{S#1}
	
	In this supplemental material (SM) we discuss the connection between the outcome of the projection operator technique using the oblique projector and non-equilibrium response theory. Additionally, we provide details on the simulation model, the reconstruction of the generalized Langevin equation (GLE), and the Markovian embedding.

    \section{Non-equilibrium response theory}

    The first fluctuation-dissipation relation (1FDR) connecting the response function $\chi(t)$ with the velocity auto-correlation function $C^V(t)$ is one of the cornerstones of statistical physics since it combines equilibrium correlation functions with non-equilibrium transport. Here, we derive a non-equilibrium generalization of the 1FDR, following the lines of Ref.~\cite{franosch2026fundamental}.

We start with the stationary dynamics of an observable $V$ in the phase-space $\Gamma$ using the stochastic operator $\mathcal{L}(t)$ and the time-evolution operator introduced in the main manuscript,
\begin{equation}
 V(\Gamma,t) = \mathcal{U}_\mathcal{L}(t,\tau) V(\Gamma,\tau).
\end{equation}
We assume that the system is for $t< \tau$ in a stationary state described by the stationary probability distribution $\rho_\text{st}(\Gamma)$. The distribution $\rho_\text{st}(\Gamma)$ therefore fulfills $\frac{\partial}{\partial t} \rho_\text{st}(\Gamma) = \mathcal{L}^*(t) \rho_\text{st}(\Gamma) = 0$ with the adjoint  operator $\mathcal{L}^*$, defined as $\langle V \mathcal{L}^*(t) U \rangle = \langle \mathcal{L}(t) V  U \rangle$.  We then introduce a perturbation, $\mathcal{L}_\delta(t) = \mathcal{L}(t) + \delta \mathcal{L}(t)$, which acts onto the variable $U$ as $\delta \mathcal{L}(t) = f(t) \frac{\partial }{\partial U}$. We therefore find
\begin{equation}
\dot{U}(t) = \mathcal{L}(t) U(t) + f(t).
\end{equation}
The time-dependent function $f(t)$ can therefore be seen as a generalized force which drives the system away from the stationary state described by $\rho_\text{st}(\Gamma)$, i.e., $f(t) = 0$ for all times $t < \tau$. Finally, we define the non-stationary average,
\begin{equation}
\langle V(t) \rangle_f = \int \diff\Gamma V \rho(\Gamma,t),
\end{equation}
using the time-dependent probability distribution in the phase space $\Gamma$ which fulfills
\begin{align}
\frac{\diff}{\diff t} \rho(t) = \frac{\partial}{\partial t} \rho(t) - \big(\mathcal{L}^*(t) + \delta \mathcal{L}(t) \big)  \rho(t) = 0,
\end{align}
due to the conservation of probability.

Solving this equation in full generality is not possible. We therefore expand $\rho(t) = \rho_\text{st} + \delta \rho(t)$ in linear order in the perturbation. We find,
\begin{align}
\frac{\partial}{\partial t} \delta \rho(t) &=  \delta \mathcal{L}(t)   \rho_\text{st} + \mathcal{L}^* \delta \rho(t)\\
&=  f(t) \frac{\partial }{\partial U} \rho_\text{st} + \mathcal{L}^* \delta \rho(t)\\
&=  f(t) \rho_\text{st} \frac{\partial }{\partial U} \ln \rho_\text{st} + \mathcal{L}^* \delta \rho(t)\\
&= f(t) J_U \rho_\text{st} + \mathcal{L}^* \delta \rho(t)
\end{align}
where we have introduced the dissipative flux $J_U=\frac{\partial }{\partial U} \ln \rho_\text{st}.$
Using the variation of constants we can therefore derive for the dynamics of $\delta \rho(t),$
\begin{align}
\delta \rho(t)  = \rho_\text{st} \int_{\tau}^{t} \diff s f(s) \mathcal{U}_{\mathcal{L}^*}(t,s) J_U,
\end{align}
where we have used the fact that $\mathcal{L}^*(t) (\rho_\text{st} J_U) = J_U \mathcal{L}^*(t) \rho_\text{st} + \rho_\text{st} \mathcal{L}^*(t) J_U = \rho_\text{st} \mathcal{L}^*(t) J_U.$
Multiplying with $V$ and averaging over the phase-space density thus yields (assuming that the stationary average $\langle V(t) \rangle_0 = 0 $),
\begin{align}
\langle V(t) \rangle_f &=  \int_{\tau}^{t} \diff s f(s) \langle V \mathcal{U}_{\mathcal{L}^*}(t,s) J_U \rangle\\
& = \int_{\tau}^{t} \diff s f(s) \langle \mathcal{U}_\mathcal{L}(t,s) V  J_U \rangle\\
&=  \int_{\tau}^{t} \diff s f(s) \langle V(t-s)  J_U \rangle.
\end{align}
Here, we have used the definition of the adjoint operator introduced above and assumed that the dynamics described by the time-evolution operator $\mathcal{U}_\mathcal{L}$ is stationary. Clearly, we can identify the response function
\begin{align}
\chi_{VU}(t) &=  \langle V(t)  J_U \rangle\\
&=\langle V(t)  \frac{\partial }{\partial U} \ln \rho_\text{st} \rangle.
\end{align}
This result is reminiscent of similar derivations, e.g., in Ref.~\cite{Vulpiani2019_fdr} using a slightly different derivation. If we insert $V=U=v_0$, we finally have
\begin{align}
\chi(t) =\langle v_0(t)  \frac{\partial }{\partial v_0} \ln \rho_\text{st} \rangle,
\end{align}
which is identical to the response function identified in the main manuscript using the oblique projector $\mathcal{P}_*.$ This derivation therefore connects the projection operator formalism with non-equilibrium linear response and shows that the memory kernel $\KS$ indeed correctly describes the dissipative response to a perturbation.

    \section{Passive tracer in a bath of chiral active Brownian particles}

    As described in the main text we adapt the model introduced in Ref.~\cite{Wijland2025PRL} with only minor changes. The first change is the inclusion of translational diffusion $D_t$ to increase the kinetic temperature of the tracer at high chiral frequency $\omega$ and generally increase diffusivity to improve mixing, in particular, for the simulations with external pulling force. The second change is the usage of a linear interaction force between the passive tracer and the active particles.

    \subsection{Chiral active Brownian particles}

    The position $\bm{{r}}_i$ of the chiral active Brownian particle with index $i$ follows the time evolution,
    \begin{equation}
        \gamma \bm{\dot{r}}_i = \bm{F}_i + f_0 \bm{u}_i + \sqrt{2 D_t} W^t_i(t), \quad \dot{\theta}_i = \omega + \sqrt{2 D_r} W^r_i(t),
    \end{equation}
    with the friction coefficient $\gamma=1$, conservative forces $\bm{F}_i$, active force $f_0$ acting in the direction $\bm{u}_i = [\cos(\theta_i),\sin(\theta_i)]^T$ of the orientation $\theta_i$, translational diffusion coefficient $D_t=1$, chiral frequency $\omega$, rotational diffusion coefficient $D_r=1,$ and uncorrelated, Gaussian white noise, $W^t_i(t)$ and $W^r_i(t)$. Throughout this work, we set the active persistence length $\ell_p = f_0 / D_r \gamma = 10$ and vary the gyroradius $\ell_g = f_0 / |\omega| \gamma \in [1,5]$ by changing $\omega \in [2,10].$ We use the Euler–Maruyama method for discretization with time step $\Delta t = 10^{-3}.$ We simulate a box of size $L_x= L_y = 20$ with 250 chiral active Brownian particles. Clearly, this will lead to finite-size effects, but since we are here mainly interested in modeling the coarse-grained dynamics of the tracer, these finite-size effects will similarly translate into the coarse-grained model. Therefore, finite-size effects will not impact the comparison of the oblique/Mori GLE with the original microscopic simulations. 

    \subsection{Passive tracer}

    The passive particle with position $\bm{R}$, momentum $\bm{P}$ and mass $m=1$ follows Hamiltonian dynamics,
    \begin{equation}
        m \bm{\dot{R}} = \bm{P}, \quad \bm{\dot{P}} = \bm{F} + \bm{F}_\text{ext},
    \end{equation}
    with reciprocal conservative force $\bm{F} = - \sum_i F_i$ and external force $\bm{F}_\text{ext}$. These equations are discretized using the velocity Verlet scheme. First, we perform $N_\text{eq} = 10^4$ equilibration steps and subsequently sample the system sufficiently long to obtain statistical errors smaller than typical data point sizes. The longest simulations are required for Fig.~4, which include $N_\text{sim} = 3 \cdot 10^8$ simulation steps. 

    \subsection{Active-passive coupling and external force}

    The active Brownian particles do not interact with each other but only with the passive tracer particle. The conservative force acting between active and passive particles is defined as,
    \begin{equation}
        \bm{F}_i =
        \begin{cases}
         k_\text{int} ( |\Delta \bm{r}_i| - r ) \Delta \bm{r}_i / |\Delta \bm{r}_i| , \quad &|\Delta \bm{r}_i| < r\\
         0, &\text{otherwise}
        \end{cases}
    \end{equation}
    with $\Delta \bm{r}_i = \bm{R} - \bm{r}_i .$ We choose $k_\text{int} = 10$ and $r = 2.5 = 0.25 \ell_p.$

    We include for Fig.~3 a harmonic external force $\bm{F}_\text{ext} = - k \bm{R}$ with harmonic strength $k$ and for Fig.~4 a constant external pulling force $\bm{F}_\text{ext} = [F_x,0]$. We choose $F_x = 0.5$ which is well within the linear response regime.

    \subsection{Response function}

    A key quantity identified in this work is the response function, $\chi(t) =\langle v_0(t)  \frac{\partial }{\partial v_0} \ln \rho_\text{st}  \rangle = \langle \frac{\partial }{\partial v_0}  v_0(t) \rangle. $ This is the average velocity $v_0(t)$ at time $t$ after an initial perturbation of strength $v_0$ at time $t=0.$

    To measure this quantity in MD and GLE simulations, we perform a standard unperturbed simulation. After a certain time $t_k$ we create a replica of the system with the only difference that the velocity of the tracer is increased by $\Delta \bm{V} = [\Delta V_x,0]$ with $\Delta V_x = 0.5.$ The response function can then be estimated as $\chi_{\alpha x}(t) = N_p^{-1} \sum_k (V_{p,\alpha} - V_{u,\alpha} ) / \Delta V_x. $ Here, $N_p = 10^6$ ($N_p =3 \cdot 10^6$ for $\omega=2$) is the number of independent perturbation runs, $V_{p,\alpha}$ is the dimension $\alpha$ of the perturbed trajectory and $V_{u,\alpha}$ of the unperturbed trajectory \cite{jung2021fdr,jung2026_nonlinear}. The subtraction of the two trajectories reduces statistical errors. Calculating these response functions with high statistical precision is by far the computational bottleneck of this study.

    \section{Reconstruction of memory kernels and fluctuating force}

While the reconstruction of the memory kernel is different for the Mori and the oblique GLE, the determination of the fluctuating force uses the same technique.

\subsection{Reconstruction of the Mori kernel}

The reconstruction of the Mori kernel follows the standard literature based on the inverse Volterra technique \cite{SHIN2010316}. The reconstruction uses the Volterra equation of second kind to determine the memory kernel from the force autocorrelation $C_{\alpha \beta}^F(t)=\langle F_\alpha(t) F_\beta(0) \rangle$, the force-velocity correlation, $C_{\alpha \beta}^{FV}(t)=\langle F_\alpha(t) V_\beta(0) \rangle$ and the velocity autocorrelation function $C_{\alpha \beta}^V(t)$,
\begin{equation}
    m^{-2} C_{\alpha \beta}^F(t) = -m^{-1} \OmegaM_{\alpha\gamma}  C_{\gamma \beta}^{FV}(t)+ \KM_{\alpha\gamma}(t) C_{\gamma \beta}^V(0)  + m^{-1} \int_0^t \diff s \KM_{\alpha\gamma}(t-s) C_{\gamma \beta}^{FV}(s).
\end{equation}
The only difference to Ref.~\cite{SHIN2010316} is that our correlation functions are matrix valued which thus also translates to the memory kernel. The above equation can be discretized in time and inverted,
\begin{align}
 \KM_{\alpha\beta}(0) &= \Big ( m^{-2} C_{\alpha \gamma}^F(0) + m^{-1} \OmegaM_{\alpha\delta} C_{\delta \gamma}^{FV}(0) \Big) C^{-1}_{\gamma \beta}\\
   \KM_{\alpha\beta}(i \Delta t) &=  \Big ( m^{-2} C_{\alpha \gamma}^F(i \Delta t) + m^{-1} \OmegaM_{\alpha\delta} C_{\delta \gamma}^{FV}(i \Delta t) - \Delta t m^{-1}  \sum_{j=0}^{i-1} w_j \KM_{\alpha\gamma}(j \Delta t) C_{\delta \gamma}^{FV}( (i -j) \Delta t)  \Big) \tilde{C}^{-1}_{\gamma \beta}
   \label{eq:Volterra_it}
\end{align}
with the weight factor $w_j = 1$ for $j>0$ and $w_0 = 1/2.$ We have also introduced $C_{\alpha \beta}^{-1}$ as the inverse of the matrix $C^V_{\alpha \beta}(0)$ and $\tilde{C}_{\alpha \beta}^{-1}$ as the inverse of the matrix $C^V_{\alpha \beta}(0) + \Delta t  m^{-1} C_{\alpha \beta}^{FV}(0). $

\subsection{Reconstruction of the oblique kernel}

As described in the main text, the oblique kernel is connected to the response function $\chi(t)$ and not to the correlation function $C^V(t)$. This leads to two difficulties compared to the Mori reconstruction. (i) The response $\chi(t)$ is more difficult to sample with high statistical accuracy since it requires a large number of independent response simulations. (ii) The reconstruction based on the Volterra equation of second kind uses the fact that the second derivative of the velocity autocorrelation functions equals the (negative) force autocorrelation function, which can be sampled with much higher accuracy than calculating numerically the second derivative. This advantage cannot be used for the response function since a similar correspondence as for correlation functions does not exist. 

The reconstruction of the oblique kernel is therefore performed in three steps. First, we fit the data for the response function $\chi(t)$ as preclusively as possible. Second, we use this fit to analytically extract the first and second derivative of the response function. Third, we use Eq.~(\ref{eq:Volterra_it}) to reconstruct the memory kernel.

For short times, the response function is best fitted using a polynomial. For longer times, it is much more precise to use oscillating exponentials. Therefore, we combine both regimes by using a smooth quintile step function.

For $t < t_1$ we fit $\chi(t)$ using a polynomial $f_p(t)$ of maximal order 6 without the linear order such that the derivative at $t=0$ is zero. Additionally, we fit the whole curve using the function $f_e(t)= \sum_{k=0}^{N_e} a_k \exp(-\gamma_k t) \cos(\omega_k t + \alpha_k).$ The final function representing $\chi(t)$ is then $f(t) = (1-w_q(t)) f_p(t) + w_q(t) f_e(t)  $, with,
\begin{equation}
    w_q(t) = \begin{cases}
        6 t^5 - 15 t^4 + 10t^3, \quad t < t_1\\
        1, \quad t \geq t_1
    \end{cases}
\end{equation}
The function $f(t)$ can be analytically differentiated, yielding $\chi^\prime(t)$ and $\chi^{\prime\prime}(t).$ This fitting procedure is done independently for the diagonal and the off-diagonal component.

Finally, we then use the Volterra equation of second kind to derive an inverse algorithm for the memory kernel,
\begin{align}
 \KS_{\alpha\beta}(0) &= - \chi^{\prime\prime}_{\alpha \gamma}(0) + \OmegaS_{\alpha\delta} \chi^\prime_{\delta \gamma}(0) \\
   \KS_{\alpha\beta}(i \Delta t) &=  \Big (  \chi^{\prime\prime}_{\alpha \gamma}(i \Delta t) +  \OmegaS{\alpha\delta} \chi^\prime_{\delta \gamma}(i \Delta t) - \Delta t   \sum_{j=0}^{i-1} w_j \KS_{\alpha\gamma}(j \Delta t) \chi^\prime_{\delta \gamma}( (i -j) \Delta t)  \Big) \tilde{\chi}^{-1}_{\gamma \beta},
   \label{eq:Volterra_it_chi}
\end{align}
where we have introduced $\tilde{\chi}_{\alpha \beta}^{-1}$ as the inverse of the matrix $\delta_{\alpha \beta} + \Delta t  \chi^\prime_{\alpha \beta}(0). $

\subsection{Reconstruction of the fluctuating force}

Having reconstructed the memory kernels $\KM$ and $\KS$, we can reconstruct the fluctuating forces $\etaM$ and $\etaS$, respectively. For this, we perform a standard MD simulation and calculate the force $\bm{F}(t)$ acting on the passive tracer at time $t$. This allows us to determine the fluctuating force 
\begin{equation}
    \eta(t) =  m^{-1}\bm{F}(t) - \Omega \bm{V}(t) + \int_{-\infty}^t \diff s K(t-s) \bm{V}(s).
\end{equation}
All terms on the right hand side are known since we have reconstructed $\Omega$ and $K(t)$ using either the Mori or oblique GLE and $\bm{V}(t)$ is just the velocity of the tracer $\bm{V}(t) = m^{-1}\bm{P}(t)$, which we determine in the MD simulation using the velocity Verlet algorithm, as described above.
    
	\section{Markovian embedding}

    To integrate both the Mori and oblique GLE, we employ a Markovian embedding in which the non-Markovian GLE is replaced by a Markovian auxiliary variable expansion \cite{ceriotti2010colored}. Different from previous approaches designed for equilibrium dynamics, we use different expansions for the memory and the fluctuating force, since the oblique GLE does not fulfill the usual second fluctuation-dissipation relation. Additionally, we model a two-dimensional stochastic process with non-trivial cross correlations which requires a more complex auxiliary expansion as detailed below.

    \subsection{Kernel embedding}

     First, we introduce an auxiliary variable system with $2 N^K_k$ additional variables to describe the off-diagonal component of the memory $K_{o}$. The equation of motion of variable $k$ is given by,
    \begin{align}
      \frac{\diff}{\diff t}  v^{K_{o}}_{1,k}(t) &= - \gamma^{K_{o}}_k v^{K_{o}}_{1,k} + \omega_k^{K_{o}} v^{K_{o}}_{2,k} - A^{K_{o}}_k V_1\\
      \frac{\diff}{\diff t}  v^{K_{o}}_{2,k}(t) &= - \gamma^{K_{o}}_k v^{K_{o}}_{2,k} - \omega_k^{K_{o}} v^{K_{o}}_{1,k} - A^{K_{o}}_k V_2.
    \end{align}
    Here, we have introduced the velocity of the tracer $\bm{V}=(V_1,V_2).$ Note, that there are no white noise contribution as in usual Markovian embedding since these variables only represent the systematic memory contribution. Second, we introduce additional $4 N^K_k$ auxiliary variables to describe the remaining diagonal components of the kernel $K_{d}$. Here, the equations of motion are given by,
        \begin{align}
      \frac{\diff}{\diff t}  v^{K_{d}}_{1,k}(t) &= - \gamma^{K_{d}}_k v^{K_{d}}_{1,k} + \omega_k^{K_{d}} v^{K_{d}}_{2,k} - \delta_{\alpha 1} A^{K_{d}}_k V_1\\
      \frac{\diff}{\diff t}  v^{K_{d}}_{2,k}(t) &= - \gamma^{K_{d}}_k v^{K_{d}}_{2,k} - \omega_k^{K_{d}} v^{K_{d}}_{1,k}  - \delta_{\alpha 2}A^{K_{d}}_k V_2.
    \end{align}
    Different from above, these variables therefore only couple to $V_1$ or $V_2$ and not to both dimensions simultaneously. 
    
    \subsection{Fluctuating force embedding}

    The fluctuating force is modeled in similar fashion as the kernel, just that we do not have to differentiate between the diagonal component, $\eta_d$,  and the off-diagonal component, $\eta_o$,
        \begin{align}
      \frac{\diff}{\diff t}  v^{\eta_{o/d}}_{1,k}(t) &= - \gamma^{\eta_{o/d}}_k v^{\eta_{o/d}}_{1,k} + \omega_k^{\eta_{o/d}} v^{\eta_{o/d}}_{2,k}  + \sqrt{2 \gamma^{\eta_{o/d}}_k} W^{\eta_{o/d}}_{1,k}(t)\\\
      \frac{\diff}{\diff t}  v^{\eta_{o/d}}_{2,k}(t) &= - \gamma^{\eta_{o/d}}_k v^{\eta_{o/d}}_{2,k} - \omega_k^{\eta_{o/d}} v^{\eta_{o/d}}_{1,k} + \sqrt{2 \gamma^{\eta_{o/d}}_k} W^{\eta_{o/d}}_{2,k}(t).
    \end{align}
    Here, $W^{\eta_{o/d}}_{j,k}(t)$ is uncorrelated Gaussian white noise. Different from above, there is no coupling back to the velocity of the tracer, instead, there is a white noise contribution. Therefore, the contribution of $v^{\eta_{o/d}}_{j,k}(t)$ to the dynamics of the coarse-grained tracer is purely stochastic. 

    \subsection{Combined Markovian embedding and effective GLE}

    Combining the above equations, we simulate the dynamics of the passive tracer using the following equation,
    \begin{align}
        \frac{\diff}{\diff t} {V}_\alpha(t) = \Omega_{\alpha \beta} {V}_\beta(t)+  \sum_{k=1}^{N_k^K} \Big( A^{K_{o}}_k  v^{K_{o}}_{\alpha,k}(t) + \tilde{A}^{K_{d}}_k v^{K_{d}}_{\alpha,k}(t) \Big) + \sum_{k=1}^{N_k^\eta} \Big( A^{\eta_{o}}_k  v^{\eta_{o}}_{\alpha,k}(t) + A^{\eta_{d}}_k v^{\eta_{d}}_{\alpha,k}(t) \Big).
    \end{align}
    If we integrate out the auxiliary variables we find a GLE with the following kernel $K_{\alpha \beta}(t)$ and fluctuating force correlation function $C^\eta_{\alpha _\beta},$
    \begin{align}
        K_{\alpha \beta}(t) &=  (1 - \delta_{\alpha \beta})  {A^{K_{o}}_k}^2 e^{-\gamma^{K_{o}}_k t} \sin(\omega_k^{K_{o}} t) + \delta_{\alpha \beta}\Big( {A^{K_{o}}_k}^2 e^{-\gamma^{K_{o}}_k t} \cos(\omega_k^{K_{o}} t) + A^{K_{d}}_k  \tilde{A}^{K_{d}}_k e^{-\gamma^{K_{d}}_k t} \cos(\omega_k^{K_{d}} t)  \Big)\\
        C^\eta_{\alpha \beta}(t) &=  (1 - \delta_{\alpha \beta})  {A^{\eta_{o}}_k}^2 e^{-\gamma^{\eta_{o}}_k t} \sin(\omega_k^{\eta_{o}} t) + \delta_{\alpha \beta}\Big( {A^{\eta_{o}}_k}^2 e^{-\gamma^{\eta_{o}}_k t} \cos(\omega_k^{\eta_{o}} t) + {A^{\eta_{d}}_k}^2  e^{-\gamma^{\eta_{d}}_k t} \cos(\omega_k^{\eta_{d}} t)  \Big).
    \end{align}
    While both equations are very similar, there is one major difference. While the diagonal memory kernel can be negative, even at $t=0$, the function $C^\eta_{\alpha \beta}(t)$ must stay positive semi-definite, which is an inherent constraint of correlation functions.

    \subsection{Determination of the fitting parameters $\gamma$, $\omega$ and $A$}

    We first fit the off-diagonal components of both the memory kernel, $K_{12}(t)$, and the correlation function, $C^\eta_{12}(t)$ using the function,
    \begin{equation}
        f^o(t) = \sum_k A_k^2 \exp(-\gamma_k) \sin(\omega_k t).
    \end{equation}
    The off-diagonal memory kernel will obviously also lead to contributions to the diagonal memory kernel. We therefore subtract these contributions from the original diagonal component, $\tilde{K}_{11}(t) = {K}_{11}(t) -  \sum_k A_k^2 \exp(-\gamma_k) \cos(\omega_k t)$ (similarly for the fluctuating force $\tilde{C}^\eta_{11}(t)$). Finally, we fit the remaining diagonal component of the memory  $\tilde{K}_{11}(t)$ using the function,
    \begin{equation}
        f^K(t) = \sum_k A_k \tilde{A}_k \exp(-\gamma_k) \cos(\omega_k t),
    \end{equation}
    and the fluctuating force $\tilde{C}^\eta_{11}(t)$ using the function,
        \begin{equation}
        f^\eta(t) = \sum_k A_k^2 \exp(-\gamma_k) \cos(\omega_k t).
    \end{equation}
    With this procedure we find all the missing parameters for the Markovian embedding. It should be noted that using too many modes for fitting the off-diagonal correlation function $C^\eta_{12}(t)$ can yield a negative $\tilde{C}^\eta_{11}(t)$, even at $t=0$, which can thus not be represented as a correlation function. Therefore it is essential to constrain the fitting. Differences observed between the results from microscopic simulations and the GLE mostly originate from these restricted fits which represent well the fluctuating force correlation function, but cannot reproduce each minor feature. 
    
	\bibliography{library.bib}